\documentclass[
 reprint,
 amsmath,amssymb,
 aps,
 prl,
]{revtex4-1}

\usepackage{graphicx}
\usepackage{dcolumn}
\usepackage{epstopdf}
\usepackage{bm}
\usepackage{color}
\usepackage{ulem}
\usepackage{float}
\restylefloat{table}
\begin{document}

\title{Interplay between Quantum Well Width and Interface Roughness for \\ Electron Transport Mobility in GaAs Quantum Wells}
\date{\today}

\author{D.\ Kamburov}
\author{K. W.\ Baldwin}
\author{K. W.\ West}
\author{M.\ Shayegan}
\author{L. N.\ Pfeiffer}
\affiliation{Department of Electrical Engineering, Princeton University, Princeton, New Jersey 08544, USA}

\begin{abstract}

We report transport mobility measurements for clean, two-dimensional (2D) electron systems confined to GaAs quantum wells (QWs), grown via molecular beam epitaxy, in two families of structures, a standard, symmetrically-doped GaAs set of QWs with Al$_{0.32}$Ga$_{0.68}$As barriers, and one with additional AlAs cladding surrounding the QWs. Our results indicate that the mobility in narrow QWs with no cladding is consistent with existing theoretical calculations where interface roughness effects are softened by the penetration of the electron wave function into the adjacent low barriers. In contrast, data from AlAs-clad wells show a number of samples where the 2D electron mobility is severely limited by interface roughness. These measurements across three orders of magnitude in mobility provide a road map of reachable mobilities in the growth of GaAs structures of different electron densities, well widths, and barrier heights.    

\end{abstract}


\maketitle

Two-dimensional electron systems (2DESs) in engineered quantum structures have proven to be a fruitful tool for discovering new fundamental physical phenomena, including the integer (IQHE) and the fractional quantum Hall effect (FQHE) \cite{Klitzing.1980,Tsui.1982}. Enabled by the progress in molecular beam epitaxy (MBE) technology \cite{Cho}, the GaAs system has become the benchmark for the highest material quality and has paved the way to very long carrier mean-free paths and high mobilities \cite{Pfeiffer1}. The present state of the MBE art allows precise control of a number of growth parameters, such as substrate temperature, rate, material composition, and growth interrupts, which critically affect scattering mechanisms and carrier mobilities \cite{Santos,Sajoto,Shayegan0,Kohrbruck,Masselink}. Generally, quantum wells (QWs) doped from both sides have to be narrower than triangular modulation-doped single-sided GaAlAs/GaAs QWs to avoid second subband occupation at high carrier concentrations \cite{Stormer}. In such QWs, small variations of the QW width have a profound effect on the energy eigenvalues. Thus interface roughness takes on additional importance for MBE growth of high density, high mobility 2DESs. Despite the immense progress in MBE techniques, however, GaAs structures where interface roughness dominates the carrier scattering have not been fully delineated systematically \cite{Gottinger,Luhman,Sakaki,Chang,Nag}. More specifically, the emergence of significant interface scattering has been demonstrated only in isolated cases in samples with sufficiently narrow QWs \cite{Sakaki} and for 2D systems with very low density \cite{Luhman}. The lack of systematic reports of carrier mobilities is surprising in view of the important role of interface roughness at high wave function confinement and its effect on carrier mobility. In QWs with AlAs cladding, where the carrier wave function is expected to have very little penetration in the barrier, the mobility was experimentally shown in a landmark paper by Sakaki \textit{et al}. \cite{Sakaki} to go as $\mu \propto W^{6}$. This agrees well with the concept that interface scattering is the major mobility-limiting factor \cite{Gottinger,Sakaki,Chang,Nag,Ando,Luhman}. On the other hand, in QWs with no AlAs cladding, where wave function penetration into the barriers is more significant, deviations from this trend should be expected \cite{Li}.

In this study we present results for 2DESs confined to symmetrically delta-doped, MBE-grown GaAs QWs on (001) GaAs substrates. Our work focuses on two sets of samples: (i) simple GaAs QWs symmetrically modulation-doped to a nominal electron density of $n \simeq 1.2 \times 10^{11}$ cm$^{-2}$, and (ii) 5.66 nm AlAs-clad QWs, in which the symmetrically modulation-doped GaAs QWs have a nominal electron density of $n \simeq 3.0 \times 10^{11}$ cm$^{-2}$. The 2D electrons in (i) are located 398 nm under the surface and are flanked on each side by undoped Al$_{0.32}$Ga$_{0.68}$As setback layers. The AlAs-clad samples are shallower, with the 2D electrons at 270 nm under the surface, flanked on each side by 5.66 nm AlAs cladding, varying-thickness undoped Al$_{0.32}$Ga$_{0.68}$As setback layers, and $\delta$-doped Si layers. The variation in the setback thickness in both sets of samples is necessary in order to maintain the 2DES density near either of the nominal values $n \simeq 1.2 \times 10^{11}$ cm$^{-2}$ or $n \simeq 3.0 \times 10^{11}$ cm$^{-2}$ to compensate for changes in the energy eigenvalues with QW width. The AlAs cladding steepens the QW barriers, thus limiting the electron wave function penetration outside the QWs. The QW widths range from $W=2-48$ nm for the simple wells at $n \simeq 1.2 \times 10^{11}$ cm$^{-2}$ and $W=0-29$ nm for the AlAs-clad QWs at $n \simeq 3.0 \times 10^{11}$ cm$^{-2}$. Each sample was measured in a van der Pauw configuration with annealed InSn contacts in a $^3$He refrigerator with base temperature of $T\simeq$ 0.3 K. Measurements were carried out using standard low-frequency lock-in amplifier techniques. Mobility values were extracted using the van der Pauw's method \cite{vanderPauw} in a square geometry using the resistance measurements from the four corners or flats of the samples. Density values were measured using standard transport measurements at perpendicular magnetic field up to 14 T.

\begin{figure}[t!]
\includegraphics[trim=0 0cm 0cm 0cm, clip=true, width=0.50\textwidth]{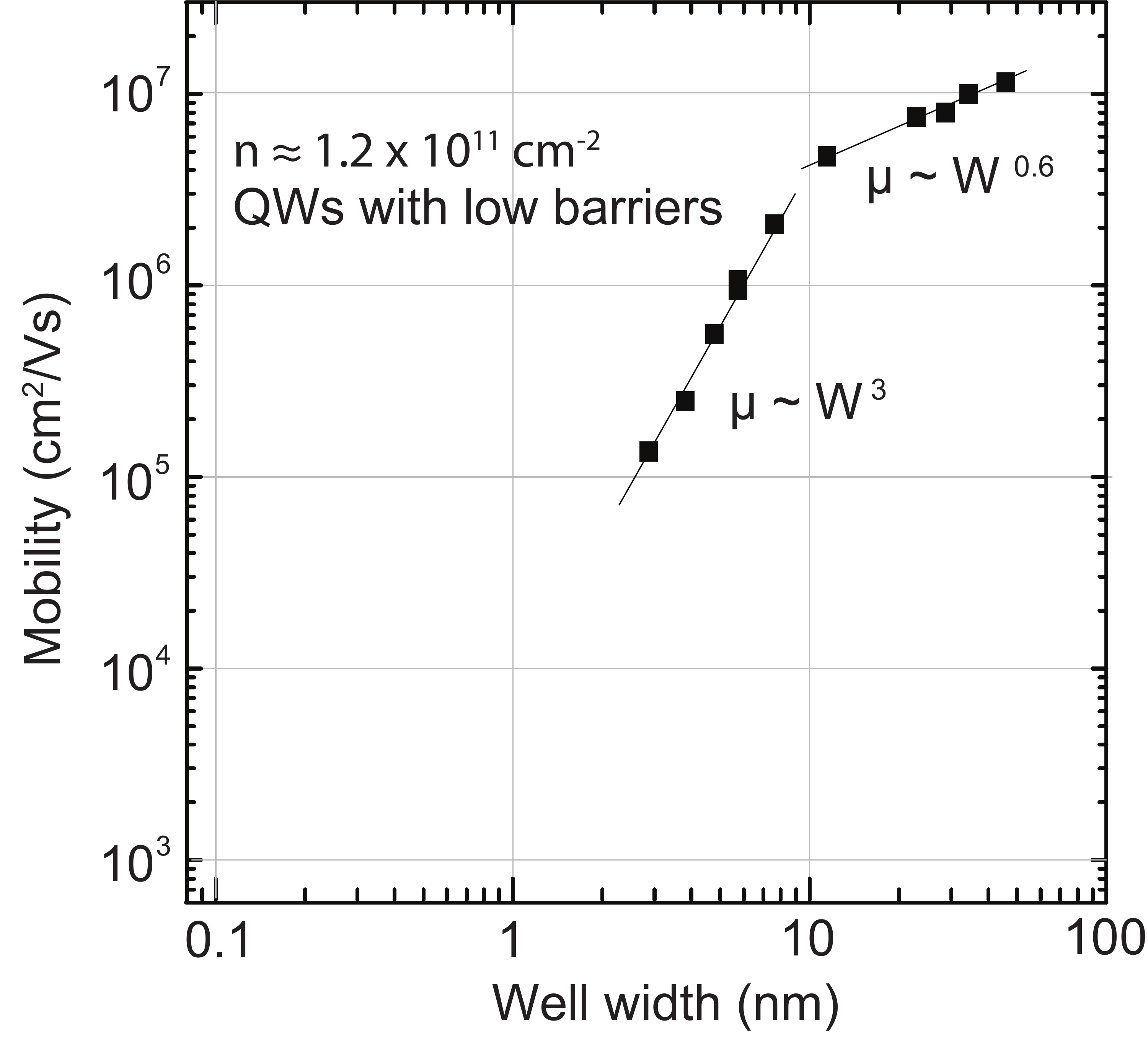}
\caption{\label{fig:Fig1} Electron mobility as a function of QW width in GaAs QWs without AlAs cladding and with 32\% AlGaAs barriers doped from both sides to an electron density of $n \simeq 1.2 \times 10^{11}$ cm$^{-2}$. We identify two regimes, for $W<10$ nm where $\mu \propto W^{3}$, and $W>10$ nm for which $\mu \propto W^{0.6}$. The straight lines are best fits to the experimental data.  }
\end{figure}

Our findings from the set of simple wells with Al$_{0.32}$Ga$_{0.68}$As barriers are summarized in Fig. 1. The measured electron mobility of each sample is plotted against its QW width. Mobilities measured in wide QWs are quite high, reaching $\simeq 12 \times 10^{6}$ cm$^2$/Vs for the 48 nm QW. Two regions of different slopes can be identified in Fig. 1. For $W>11$ nm, the mobility dependence is $\mu \propto W^{0.6}$, while with decreasing well width, mobilities degrade more quickly as approximated by $\mu \propto W^{3}$. The gradual decrease of $\mu$ in narrow QWs is consistent with numerical calculations based on the theoretical model of Refs. \cite{Li,Sakaki}, in which the QW barrier is treated as \textit{finite}, fully taking into account the penetration of the electron wave function into the barriers. Effectively, the absence of AlAs cladding allows the electron wave function to leak outside the QW and leads to a reduced significance of the interface roughness and thus to an overall higher mobility. Our data qualitatively match the behavior of the theoretical model in Ref. \cite{Li}. 

While the data in Fig.\;1 characterize the general case in which the electron wave function penetrates into the barriers, a more interesting scenario can be achieved when electrons are tightly confined in the QWs by adding AlAs cladding. Due to the larger band gap of AlAs, the cladding acts as a steeper potential barrier that confines the electrons to a profile more closely resembling an infinite QW. In Fig. 2 we present data from a number of AlAs-clad QWs, including a structure in which the GaAs QW is removed altogether, so that the 2DES resides in an AlAs QW. The earlier experimental results by Sakaki \textit{et al.} \cite{Sakaki}, which comprised only four low mobility data points, are included for reference in Fig. 2 with open circles. 

\begin{figure}[t!]
\includegraphics[trim=0 0.0cm 0cm 0cm, clip=true, width=0.50\textwidth]{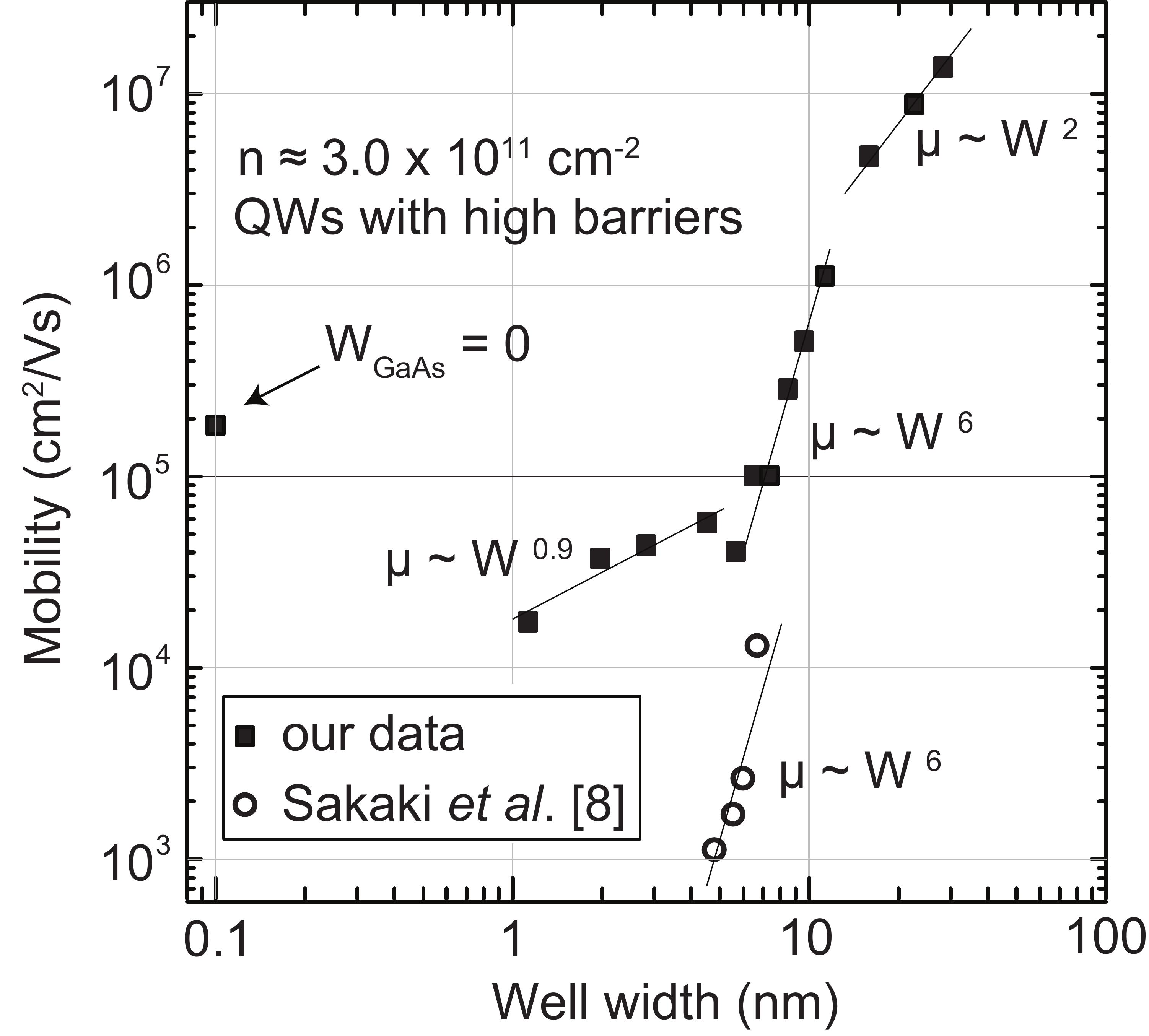}
\caption{\label{fig:Fig2} ($\blacksquare$) Electron mobility for AlAs-clad GaAs QWs doped from both sides to $n \simeq 3.0 \times 10^{11}$ cm$^{-2}$. Three regimes are prominent in the mobility curve: $W>11$ nm, where interface scattering is insignificant;  $8.65$ nm $<W<11$ nm, where interface scattering is significant and $\mu \propto W^6$; and $W<8.65$ nm, for which the width of the GaAs QW dependence is fairly flat (see text). ($\circ$) Reproduced data from Ref. \protect\cite{Sakaki}, showing the important role of interface roughness as a limiting mechanism in the electron mobility. All straight lines are best fits to the experimental data. Note that in the region $5.65\text{ nm}<W<10$ nm both data sets can be approximated by parallel lines with $\mu \propto W^6$. }
\end{figure}

Electron mobilities from the AlAs-clad QWs in Fig. 2 show a complex dependence on the GaAs QW width. In wide QWs ($W>11$ nm), where monolayer-scale interface roughness effects are expected to be less significant compared to the scale of the QW width, our experiments show $\mu \propto W^{2}$. Such behavior is not unexpected as the interface roughness should become important only in narrower QWs, where local QW width variations would severely change the local QW eigenvalue. As $W$ decreases to about 10 nm, $\mu$ is affected more significantly by the interface roughness. In the range $W=5.6-11$ nm, the mobility trend is well approximated by $\mu \propto W^{6}$, in accord with previous experiments \cite{Gottinger,Sakaki} and with slopes parallel to our data. The mobilities in our samples are almost two orders of magnitude higher, likely because we used growth temperature of 647$^{\circ}$C, while the ones from Sakaki \textit{et al.} \cite{Sakaki} were grown at 590$^{\circ}$C. For QWs narrower than $W=5.66$ nm, $\mu$ increases slightly, then decreases gradually again \cite{footnote2}. At $W_{\textbf{GaAs}}=0$, the mobility is much higher. This behavior can be qualitatively understood as follows: for narrower GaAs QW well-width ($W=4.53$ nm and lower), the tight confinement causes the wave function to substantially penetrate into the AlAs barriers and make an AlAs bilayer system \cite{Vakili} with a GaAs barrier. This is the failing point of the infinite-barrier model because the wave function effectively begins to reside more in the AlAs cladding layers and therefore relatively less in the GaAs layer. The mobility at $W=4.53$ nm increases slightly compared to the $W=5.66$ nm QW, possibly because the penetration into the AlAs cladding provides net additional room for the electron wave function. For smaller GaAs QW widths, the mobility starts to drop because the total thickness of the AlAs/GaAs/AlAs system is decreasing from $4.53$nm$+11.32$nm to $1.13$nm$+11.32$nm and the electrons are again becoming more confined. As the GaAs QW disappears altogether, $\mu$ increases, reaching values of mobility of a simple 2DESs confined to AlAs QWs \cite{DePoortere}. The wave function then fully resides in an AlAs QW of $W_{\textbf{AlAs}}=11.32$ nm, i.e. twice the width of the AlAs cladding on each side of the original GaAs well \cite{footnote1}. The additional GaAs/AlAs interfaces have also disappeared and with them the extra scattering caused by the GaAs QW.

The results from the AlAs-clad QWs corroborate the previous experimental studies and theoretical models for narrow systems with $W<10$ nm with strong confinement \cite{Gottinger,Sakaki} and very low electron density \cite{Luhman}, while adding an unexpectedly rich behavior of carrier mobilities away from this regime. In particular, when the interface roughness is insignificant on the scale of $W$, $\mu$ is only slightly affected by surface scattering because the ground-state eigenvalue of the wide QW does not locally vary much. In cases when the QW is too narrow to contain the wave function, the 2DES migrates into the AlAs cladding and the system resembles an AlAs bilayer. Finally, when the GaAs is completely removed, $\mu$ improves to the point that the sample with no GaAs well has similar quality to the GaAs well with $W=7.6$ nm. This complex trend of $\mu$ as a function of the GaAs QW width is important for understanding the interplay between interface roughness and quantum confinement. However, $\mu$ is not necessarily the only relevant characteristic in the study of 2D phenomena in engineered GaAs QWs. 

\begin{figure}[t!]
\includegraphics[trim=0 0.0cm 0cm 0cm, clip=true, width=0.49\textwidth]{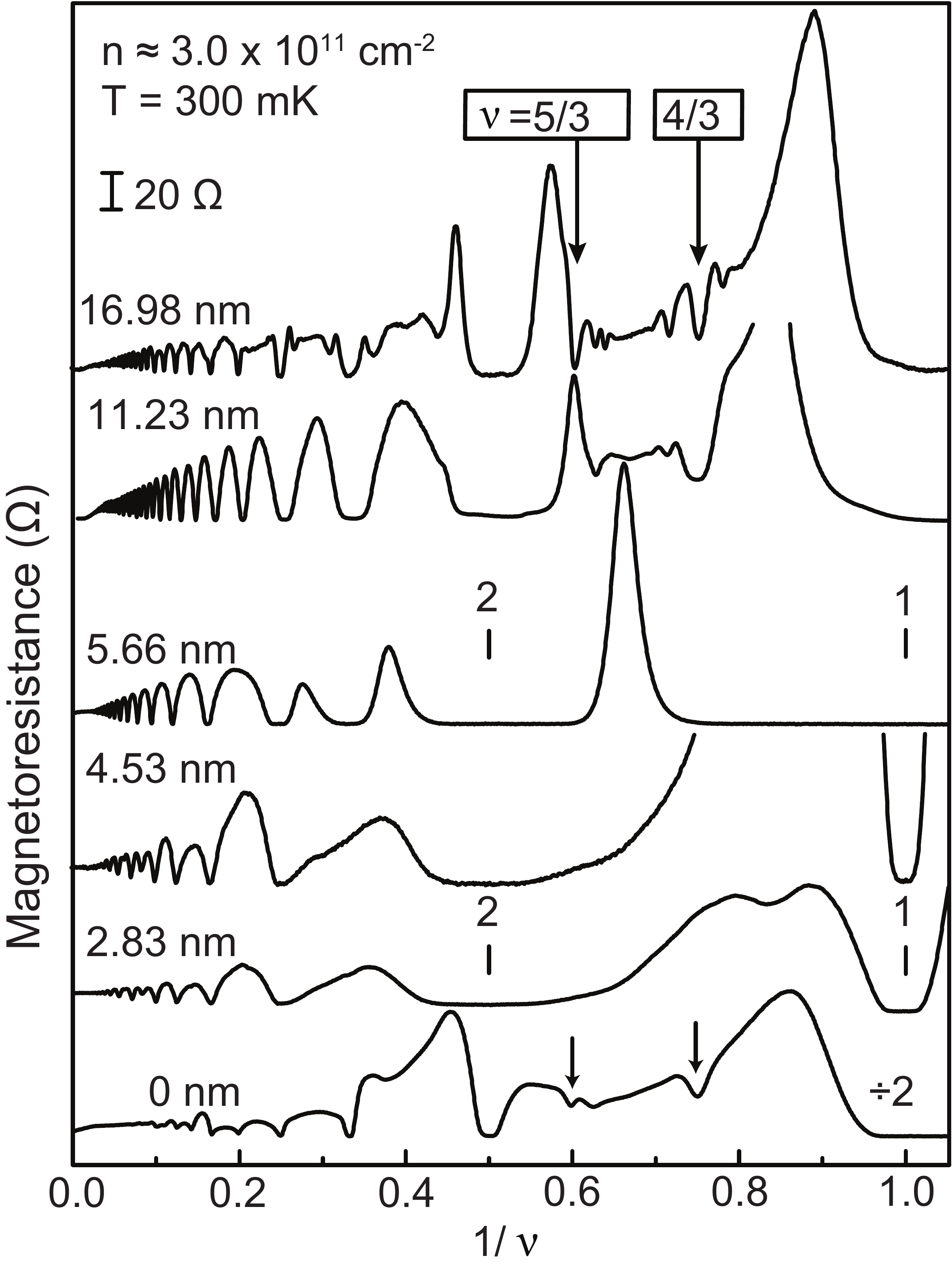}
\caption{\label{fig:Fig3} Magnetoresistance traces vs $1/\nu$, where $\nu$ is the Landau filling factor, for different AlAs-clad QWs, vertically offset for clarity. The electron density in all samples is $n \simeq 3.0 \times 10^{11}$ cm$^{-2}$.}
\end{figure}

In order to provide a more comprehensive understanding of the implications of the sample quality for a given value of $\mu$, in Fig. 3 we plot representative magnetoresistance traces taken for the AlAs-clad QWs whose mobilities are summarized in Fig. 2. The samples with $W>10$ nm have high quality and their magnetoristance traces show signatures of FQHE states, at Landau level filling factors $\nu=5/3$ and $4/3$, marked with vertical arrows. In the $W=5.66$ nm QW trace, the FQHE states disappear and the trace shows strong and wide IQHE minima. For just slightly narrower QWs, $W=4.53$ nm, the $\nu=1$ minimum gets strikingly narrower, a signature of AlAs bilayer effect \cite{Vakili}. In sharp contrast, when the GaAs QW is removed, the sample becomes a wide AlAs QW of $W_{\text{AlAs}}=11.32$ nm, and the magnetoresistance trace again shows FQHE states \cite{DePoortere}. This improvement results from removing the narrow GaAs QW as a scattering mechanism and an intervening barrier. We note that for all AlAs-clad samples with $W<5.66$ nm, we see bilayer AlAs-like characteristics, namely a weak $\nu=1$ state and absence of minima at odd fillings \cite{Vakili,Shayegan}. The magnetoresistance data from the lower density set of samples (not shown), on the other hand, lack any unusual behavior because of the deep penetration of the wave function into the low barriers and the effective larger quantum well width. 

To summarize, our data reveal: (1) a $\mu \propto W^{3}$ dependence of mobility in simple GaAs QWs with Al$_{0.32}$Ga$_{0.68}$As barriers, consistent with muted interface roughness, typical of systems with only modestly high potential barriers, (2) a $\mu \propto W^{6}$ relationship for AlAs-clad narrow GaAs QWs occurs where interface roughness is significant, but only in a narrow range of QW widths, in general agreement with previous experiments, (3) unusually rich $W$-dependence of $\mu$ in narrow AlAs-clad QWs, signaling an AlAs bilayer, and (4) much higher $\mu$ in the complete absence of a GaAs QW, when the 2D electrons reside in a clean AlAs QW. We emphasize that the our results provide a comprehensive road map of the $\mu$-dependence on QW width for a variety of structures.

\begin{acknowledgments}
We acknowledge support through the Gordon and Betty Moore Foundation (Grant GBMF4420) and NSF (DMR-1305691, ECCS-1508925, MRSEC DMR-1420541).
\end{acknowledgments}


\begin{thebibliography}{99}

\bibitem{Klitzing.1980} K. von Klitzing, G. Dorda, and M. Pepper, Phys. Rev. Lett. \textbf{45}, 494–497 (1980).
\bibitem{Tsui.1982} D. C. Tsui, H. L. Stormer, and A. C. Gossard, Phys. Rev. Lett. \textbf{48}, 1559 (1982).
\bibitem{Cho} A. Y. Cho, J. Appl. Phys. \textbf{41}, 2780, (1970).
\bibitem{Pfeiffer1} L. N. Pfeiffer, K. W. West, H. L. Stormer, and K. W. Baldwin, Appl. Phys. Lett. \textbf{55}, 1888 (1989).
\bibitem{Santos} M. Santos, T. Sajoto, A. Zrenner and M. Shayegan, Appl. Phys. Lett. \textbf{53}, 2504 (1988).
\bibitem{Sajoto} T. Sajoto, M. Santos, J. J. Heremans, M. Shayegan, M. Heiblum, M. V. Weckwerth, and U. Meirav, Appl. Phys. Lett. \textbf{54}, 840 (1989).
\bibitem{Shayegan0} M. Shayegan, V. J. Goldman, M. Santos, T. Sajoto, L. Engel, and D. C. Tsui, Appl. Phys. Lett. \textbf{53}, 2080 (1988).
\bibitem{Kohrbruck} R. Kohrbruck, S. Munnix, D. Bimberg, D. E. Mars, and J. N. Miller,
Appl. Phys. Lett. \textbf{57}, 1025 (1990).
\bibitem{Masselink} W. T. Masselink, Y. L. Sun, R. Fischer, T. J. Drummond, Y. C. Chang, M. V. Klein, and H. Morkoc¸, J. Vac. Sci. Technol. B \textbf{2}, 117 (1984).
\bibitem{Stormer} H.L. Stormer, A.C. Gossard, W. Wiegmann, Solid State Commun. \textbf{41}, 10 (1982).
\bibitem{Sakaki} H. Sakaki, T. Noda, K. Hirakawa, M. Tanaka, and T. Matsusue, Appl. Phys. Lett. \textbf{51}, 1934 (1987).
\bibitem{Gottinger} R. Gottinger, A. Gold, G. Abstreiter, G. Weimann, and W. Schlapp, Europhys. Lett., \textbf{6}, 183-188 (1988).
\bibitem{Luhman} D. R. Luhman, D. C. Tsui, L. N. Pfeiffer, and K. W. West, Appl. Phys. Lett. \textbf{91}, 072104 (2007).
\bibitem{Chang} T. Chang, L. P. Fu, F. T. Bacalzo, G. D. Gilliland, D. J. Wolford, K. K. Bajaj, A. Antonelli, R. Chen, J. Klem, and M. Hafich, J. Vac. Sci. Tech. B \textbf{13}, 1760 (1995).
\bibitem{Nag} B. R. Nag, Semicond. Sci. Technol. \textbf{19}, 162–166 (2004).
\bibitem{Ando} T. Ando, J. Phys. Soc. Jpn. \textbf{51}, 3900 (1982).
\bibitem{Li} J. M. Li, J. J. Wu, X. X. Han, Y. W. Lu, X. L. Liu, Q. S. Zhu, and Z. G. Wang, Semicond. Sci. Tech. \textbf{20}, 1207–-1212 (2005).
\bibitem{vanderPauw} L. J. van der Pauw, Philips Res. Repts. \textbf{13}, 1-9 (1958). 
\bibitem{footnote2} The mobilities measured in samples near and at $W=5.66$ nm are quite robust across multiple wafers.
\bibitem{Vakili} K. Vakili, Y. P. Shkolnikov, E. Tutuc, E. P. De Poortere, and M. Shayegan, Phys. Rev. Lett. \textbf{92}, 186404 (2004).
\bibitem{DePoortere} E. P. De Poortere, Y. P. Shkolnikov, E. Tutuc, S. J. Papadakis, M. Shayegan, E. Palm, and T. Murphy, Appl. Phys. Lett. \textbf{80}, 1583 (2002).
\bibitem{footnote1} Note that this last point in Fig. 2 is plotted as $W=0.1$ nm because of the logarithmic nature of the plot. 
\bibitem{Shayegan} M. Shayegan, E. P. De Poortere, O. Gunawan, Y. P. Shkolnikov, E. Tutuc, and K. Vakili, Phys. Stat. Sol. (b) \textbf{243}, 14 (2006).
 
 
 


\end{thebibliography}
\end{document}